\documentclass[11pt]{article}
\usepackage{amssymb,amsmath,graphicx,braket,hyperref}
\usepackage{blkarray}
\newcommand{\mLabel}[1]{\mbox{$\scriptstyle{#1}$}}

\begin{document}

\begin{titlepage}

    \begin{flushright}
      \normalsize TU--1028\\
      \today
    \end{flushright}

\vskip2.5cm
\begin{center}
\LARGE
On enhanced corrections
from quasi-degenerate states\\
to heavy quarkonium observables 
\end{center}

\vspace*{0.8cm}
\begin{center}
{\sc Y. Kiyo}$^{a}$,
{\sc G. Mishima}$^{b}$ and
{\sc Y. Sumino$^{b}$}\\[5mm]
  {\small\it $^a$ Department of Physics, Juntendo University}\\[0.1cm]
  {\small\it Inzai, Chiba 270-1695, Japan}

  {\small\it $^b$ Department of Physics, Tohoku University}\\[0.1cm]
  {\small\it Sendai, 980-8578 Japan}

\end{center}

\vspace*{2.8cm}
\begin{abstract}
\noindent
It is well known that 
in perturbation theory
existence of quasi-degenerate states
can rearrange the order counting.
For a heavy quarkonium system, naively, enhanced effects 
($l$-changing mixing effects) could
contribute already to the first-order and third-order
corrections to the wave function and the energy level, respectively,
in expansion in $\alpha_s$.
We present a formulation and note that the corresponding
(lowest-order)
corrections vanish due to absence of the
relevant off-diagonal matrix elements.
As a result,
in the quarkonium energy level and leptonic decay width, 
the enhanced effects are expected to appear, respectively, in the fifth-
and fourth-order 
corrections and beyond.
\vspace*{0.8cm}
\noindent

\end{abstract}


\vfil
\end{titlepage}

\newpage

\section{Introduction}
\label{}

Heavy quarkonium,
the bound state of a heavy quark-antiquark pair,
is a prime example of
a strongly interacting system
whose properties are
well documented in perturbative QCD.
With the advent of new theoretical framework, such as
effective field theory (EFT) and threshold expansion
technique, as well as proper treatment for
decoupling infrared degrees of freedom,
the heavy quarkonium system has become an ideal laboratory
for precision tests of predictions of perturbative QCD with
respect
to various experimental data and lattice QCD predictions.

The state-of-the-art  computational results in this field comprise
the next-to-next-to-next-to-leading order
(NNNLO) energy levels of heavy 
quarkonium~\cite{Beneke:2005hg,Penin:2005eu,Kiyo:2014uca},
the NNNLO pair-production cross section of heavy quarks 
near threshold~\cite{Beneke:2015kwa,Beneke:2016kkb},
and the leptonic decay width of $\Upsilon (1S)$ 
state~\cite{Beneke:2014qea}.
These calculations utilize the modern EFT,
potential-nonrelativistic QCD 
(pNRQCD)~\cite{Pineda:1997bj,Brambilla:1999xf},
for systematically organizing the perturbative expansions
in $\alpha_s$ and $v$ (velocity of heavy quarks)
in a sophisticated manner.
This EFT describes interactions of a non-relativistic
quantum mechanical system (dictated by the Schr\"odinger equation)
with ultrasoft gluons, which is organized in multipole expansion.
We can benefit from
methods and knowledge of
perturbation theory
of quantum mechanics therein.

It is widely known that
quasi-degenerate systems
need special care
in perturbation theory 
of quantum mechanics~\cite{Weinberg:2015ww},
however, thus far the relevant consideration
seems to be missing in the computation
of the aforementioned NNNLO heavy quarkonium observables.\footnote{
In the computation of the NNLO
energy levels
the corrections from quasi-degenerate states for the $n\leq 3$ states 
were explicitly
considered and found to be
absent \cite{Titard:1994id}.
}
In perturbative expansion of
the heavy quarkonium system, the leading-order
Hamiltonian is that of the Coulomb system whose energy
eigenvalues are labeled only by the principal quantum number $n$.
The first-order correction resolves the
degeneracy in the orbital angular momentum $l$,
while the second-order correction resolves the
degeneracy in the total spin  $s$ and total angular momentum $j$.
Once these features are
properly taken into account
in perturbative calculations there are enhanced contributions
which rearrange the order counting.
These are the mixing effects between different $l$ states
for the same $n$.
One finds that naively these start from the third-order corrections
to the heavy quarkonium energy levels and from the first-order
corrections to the wave functions.\footnote{
As a simple example, consider a matrix
$$
\left(
\begin{array}{cc}
0& x^2 V_2\\
x^2 V_2^* & x V_1
\end{array}
\right) .
$$
Its energy eigenvalues are given by $-x^3|V_2|^2/V_1$ and
$xV_1+x^3|V_2|^2/V_1$ up to ${\cal O}(x^3)$, and the corresponding eigenvectors
are given by $(1,-xV_2^*/V_1)$ and $(xV_2/V_1,1)$ up to ${\cal O}(x)$.
The appearance of $V_1$ in the denominator signals enhanced contributions.
}
The latter would induce 
second-order
corrections to the heavy quark threshold production cross section
(or the quarkonium leptonic decay width).
By explicit computation the relevant lowest order
off-diagonal matrix elements for these corrections vanish.
Hence, these enhanced corrections are pushed to higher orders.
We present a necessary formulation, an explicit computation
at the lowest order, and discuss further higher-order effects.

It is not our purpose to claim originality of the present work
but rather to recollect relevant information and to clarify the
basis for systematic computation.
A closely related subject is the inclusion of
transitions (mixings) between two quasi-degenerate states
given by off-diagonal matrix elements
of interaction operators 
considered in many potential-model 
calculations~\cite{Voloshin:2007dx,Segovia:2008zz,Segovia:2013wma,Segovia:2016xqb}.
However, (somewhat to our surprise) systematic order counting
in light of pNRQCD in expansion in $\alpha_s$ and $v$
has not been addressed so far.

In the case of QED,
it was already pointed out in the late 1940s and 1950s
that contributions from quasi-degenerate states
to the positronium energy levels
do not appear at and below order $\alpha ^6$
(order $\alpha^4$ relative to the LO energy levels);
see Ref.\cite{Adkins:1999hf} and references therein.
However, the situations of positronium and heavy quarkonium systems
differ in some aspects
and it is worth clarifying the latter case explicitly.
The crucial difference stems from the fact that
the degeneracy of the heavy quarkonium energy level
is lifted at $\alpha _s^3$ 
whereas the degeneracy is lifted at $\alpha ^4$ in the case of positronium.

\section{Perturbation theory for quasi-degenerate system}
\label{sec:2}

Consider the Schr$\ddot{\mathrm{o}}$dinger
equation of heavy quarkonium
\begin{align}
  \left(H^{(0)}+\sum _{i=1}^\infty \varepsilon ^i V^{(i)}
  \right)
  \ket{\Psi _{nlsj}}
  =E_{nlsj}
  \ket{\Psi _{nlsj}},
  \label{eq:schrodinger}
\end{align}
which dictates the quantum mechanical
subsystem in pNRQCD.
An expansion parameter $\varepsilon$ (corresponding to $\alpha_s$ or $v$)
is introduced\footnote{
For simplicity
we neglect the electromagnetic interaction
of quarks.
In the case of bottom quark,
numerically its effects are small even compared to the NNNLO
corrections in $\alpha_s$.
The electric charge of bottom quark $Q_b=-1/3$
plays a role of an extra suppression factor
in addition to the small
QED coupling constant $\alpha \simeq 1/137$,
as compared to, e.g., $\alpha _s(m_b)\simeq 0.23$.
},
and a unique order in $\varepsilon$ is assigned to
each potential operator $V^{(i)}$.
The definitions of $H^{(0)}, V^{(1)},\dots$ can be found, for
instance, in \cite{Kniehl:2002br,Kiyo:2014uca},
but we do not need their explicit forms
in this section.
The energy level and
the wave function are labeled with
$(n,l,s,j)$.
The operators $H^{(0)}$ and $V^{(1)}$ preserve $l$, $s$ and $j$.
Furthermore, $H^{(0)}$ and $V^{(i)}$ 
preserve $s$ and $j$ (see Sec.~\ref{s4}),
hence we suppress these two labels in the following.
($\ket{nl}=\ket{nlsj}$ represents an eigenstate of $H^{(0)}$.)

The perturbative expansion of the energy level is given by
\begin{align}
  &E_{nlsj}
  =E^{(0)}_n 
  +\varepsilon E^{(1)}_{nl}
  +\varepsilon^2 \left[ \bra{nl}V^{(2)}\ket{nl} 
  +\sum_{n'\neq n}^{}\frac{|\bra{nl}V^{(1)}\ket{n'l}|^2}{E_n^{(0)}-E_{n'}^{(0)}}\right]\nonumber\\
  &~~~~~~+\varepsilon^3 \left[ \bra{nl}V^{(3)}\ket{nl} 
  +\sum_{l'\neq l}^{}\frac{|\bra{nl}V^{(2)}\ket{nl'}|^2}{E_{nl}^{(1)}-E_{nl'}^{(1)}}
  +\sum_{n'\neq n}^{}\frac{\bra{nl}V^{(2)}\ket{n'l}
	\bra{n'l}V^{(1)}\ket{nl}}{E_n^{(0)}-E_{n'}^{(0)}}\right.\nonumber\\
	&~~~~~~~~~~~~~~+\left. \sum_{n'\neq n}^{}\frac{ \bra{nl}V^{(1)}\ket{n'l}}{E^{(0)}_n-E^{(0)}_{n'}}
	\left\{
	  \bra{n'l}V^{(2)}\ket{nl}
	  -E^{(1)}_{nl}\frac{\bra{n'l}V^{(1)}\ket{nl}}{E^{(0)}_n-E^{(0)}_{n''}}
\right. \right.
\nonumber\\ &
\left. \left. 
~~~~~~~~~~~~~~~~~~~~~~~~~~~~~~~~~~~~~~~
+\sum_{n''\neq n}^{}
	  \frac{\bra{n'l}V^{(1)}\ket{n''l}\bra{n''l}V^{(1)}\ket{nl}}{E^{(0)}_n-E^{(0)}_{n''}}
	\right\}\right]
	+\mathcal{O}(\varepsilon^4),
  \label{eq:energy}
\end{align}
where we use short-hand notations
$E^{(0)}_n\equiv \bra{nl}H^{(0)}\ket{nl}$,
$E^{(1)}_{nl}\equiv \bra{nl}V^{(1)}\ket{nl}$.
The fourth- and fifth-order corrections
will be given in eqs.\eqref{eq:energy4},\eqref{eq:energy5}.
The subscript of $E^{(0)}_n$
indicates that the leading energy eigenvalue
depends only on $n$,
and that of  $E^{(1)}_{nl}$
indicates that degeneracy in $l$
is resolved at the first-order.
The degeneracy is fully resolved at the second order.

The $\varepsilon^3$-term proportional to 
$|\bra{nl}V^{(2)}\ket{nl'}|^2$
in eq.\eqref{eq:energy}
is the main focus of this paper.
This correction has not been considered explicitly
in the previous studies \cite{Beneke:2005hg,Penin:2005eu,Kiyo:2014uca}.
Since the operator $V^{(2)}$ is accompanied by $\varepsilon^2$,
naive order counting indicates that the $|\bra{nl}V^{(2)}\ket{nl'}|^2$ term
may be order $\varepsilon^4$.
Due to the quasi-degeneracy of
the states $\ket{nl}$ and $\ket{nl'}$, however,
the denominator 
$(E_{nl}^{(1)}-E_{nl'}^{(1)})$
compensates one $\varepsilon$,
rendering the term to be order $\varepsilon^3$.

The perturbative expansion of the wave function is given by
\begin{align}
  \ket{\Psi_{nlsj}}&=
  \ket{nlsj}
  +\sum _{i=1}^\infty 
  \varepsilon^i
  \left[
    \sum_{l'\neq l}^{}\ket{nl'sj}
    \frac{c^{(i)}_{nl';nl}}{E^{(1)}_{nl}-E^{(1)}_{nl'}}
    +\sum_{n'\neq n,\ l'}^{}\ket{n'l'sj}
    \frac{d^{(i)}_{n'l';nl}}{E^{(0)}_{n}-E^{(0)}_{n'}}
  \right] ,
  \label{eq:wavefunc}
\end{align}
where
$\ket{\Psi_{nlsj}}$ is normalized as $\braket{nlsj|\Psi_{nlsj}}=1$.
The coefficients are given by 
\begin{align}
c^{(1)}_{nl';nl}
=&\bra{nl'}V^{(2)}\ket{nl}
,\qquad \qquad
d^{(1)}_{n'l';nl}
=\bra{n'l'}V^{(1)}\ket{nl},
\label{eq:c1d1}\\
  c^{(2)}_{nl';nl}
  =&\bra{nl'}V^{(3)}\ket{nl}
  +\bra{nl'}V^{(2)}\ket{nl'}
  \frac{c^{(1)}_{nl';nl}}
  {E^{(1)}_{nl}-E^{(1)}_{nl'}}\nonumber\\
  &+\sum_{i=1}^{2}
  \sum_{n''\neq n,\ l''}^{}
  \bra{nl'}V^{(3-i)}\ket{n''l''}
  \frac{d^{(i)}_{n''l'';nl}}
  {E^{(0)}_{n}-E^{(0)}_{n''}}
  -E^{(2)}_{nl}
  \frac{c^{(1)}_{nl';nl}}
  {E^{(1)}_{nl}-E^{(1)}_{nl'}}, \label{eq:c2}\\
  d^{(2)}_{n'l';nl}
  =&\bra{n'l'}V^{(2)}\ket{nl}
  +\bra{n'l'}V^{(1)}\ket{nl'}
  \frac{c^{(1)}_{nl';nl}}
  {E^{(1)}_{nl}-E^{(1)}_{nl'}}\nonumber\\
  &+\sum_{n''\neq n,\ l''}^{}
  \bra{n'l'}V^{(1)}\ket{n''l''}
  \frac{d^{(1)}_{n''l'';nl}}
  {E^{(0)}_{n}-E^{(0)}_{n''}}
  -E^{(1)}_{nl}
  \frac{d^{(1)}_{n'l';nl}}
  {E^{(0)}_{n}-E^{(0)}_{n'}},
\\
    c^{(3)}_{nl';nl}
  =&\bra{nl'}V^{(4)}\ket{nl}
  +\sum_{i=1}^{2}
  \sum_{l''\neq l}^{}
  \bra{nl'}V^{(4-i)}\ket{nl''}
  \frac{c^{(i)}_{nl'';nl}}
  {E^{(1)}_{nl}-E^{(1)}_{nl''}}\nonumber\\
  &+\sum_{i=1}^{3}
  \sum_{n''\neq n,\ l''}^{}
  \bra{nl'}V^{(4-i)}\ket{n''l''}
  \frac{d^{(i)}_{n''l'';nl}}
  {E^{(0)}_{n}-E^{(0)}_{n''}}
  -\sum_{i=1}^{2}
  E^{(4-i)}_{nl}
  \frac{c^{(i)}_{nl';nl}}
  {E^{(1)}_{nl}-E^{(1)}_{nl'}},\\
  d^{(3)}_{n'l';nl}
  =&\bra{n'l'}V^{(3)}\ket{nl}
  +\sum_{i=1}^{2}
  \sum_{l''\neq l}^{}
  \bra{n'l'}V^{(3-i)}\ket{nl''}
  \frac{c^{(i)}_{nl'';nl}}
  {E^{(1)}_{nl}-E^{(1)}_{nl''}}\nonumber\\
  &+\sum_{i=1}^{2}
  \sum_{n''\neq n,\ l''}^{}
  \bra{n'l'}V^{(3-i)}\ket{n''l''}
  \frac{d^{(i)}_{n''l'';nl}}
  {E^{(0)}_{n}-E^{(0)}_{n''}}
  -\sum_{i=1}^{2}
  E^{(3-i)}_{nl}
  \frac{d^{(i)}_{n'l';nl}}
  {E^{(0)}_{n}-E^{(0)}_{n'}}.
\label{eq:d3}
\end{align}
Here, $E_{nl}^{(k)}$ denotes the coefficient of $\varepsilon^k$
of $E_{nlsj}$ [c.f., eq.(\ref{eq:energy})], 
and it is understood that 
$c^{(i)}_{nl';nl}=0$ 
if $l'= l$. 
As can be seen, 
an enhanced contribution proportional to
$c^{(1)}_{nl';nl}/
  \Bigl[ E^{(1)}_{nl}-E^{(1)}_{nl'}\Bigr]$
appears already at the first order for the wave function.
A derivation of the perturbative expansions of $E_{nlsj}$ and
$\ket{\Psi_{nlsj}}$ is given in the appendix.

Physical quantities, such as
the quark pair production cross section near threshold in
$e^+e^-$ collisions\footnote{
These corrections can make sense only in the close vicinity of 
distinct quasi-degenerate resonance
peaks.  Otherwise the enhancement from the small denominators are lost,
for instance, by smearing due to the resonance widths.
}
and the quarkonium leptonic decay width,
are proportional to the absolute square of
the wave function at the origin, and
the enhanced corrections to these observables
arise from the second order.
This is because only the $S$-wave ($l=0$) wave functions have non-vanishing 
values at the origin, and since 
the enhanced corrections should connect different $l$s,
namely they should be proportional to $|\bra{n,l=0}V^{(2)}\ket{n,l'\neq 0}|^2$.

So far we have implicitly assumed that
$\bra{nl}V^{(2)}\ket{nl'}\neq 0$ for $l'\neq l$,
but if this matrix element vanishes
for some reasons,
the enhanced corrections from quasi-degenerate states
do not appear
at least up to the fourth order
in the energy level, as well as 
up to the third order in the quark pair production cross section
(or quarkonium leptonic decay width).\footnote{
There is also a contribution from
the $D$-wave production (or decay) via the higher-dimensional
local current operator.  
In the case $\bra{nl}V^{(2)}\ket{nl'}= 0$,
contribution of enhanced corrections 
through such an operator also starts from the fourth-order correction.
}

\section{Vanishing off-diagonal matrix elements of \boldmath{$V^{(2)}$}}

In this section we evaluate
the matrix element $\bra{nl}V^{(2)}\ket{nl'}$
explicitly and show that it vanishes if $l'\neq l$.\footnote{
The vanishing of the matrix elements relies on the property of radial wave functions 
of Coulomb system $H_0=p^2/m -C_F \alpha_s/r$.
If one takes another wave function, for instance one in phenomenological 
potential models  the matrix elements can have non-zero values, while 
it violates rigid order counting of pNRQCD and bring yet higher-order effects
considered here  into the computation of the matrix elements. 
}
The operator in $V^{(2)}$
which can have non-vanishing off-diagonal matrix elements
for different $l$s is the
so-called tensor operator
\begin{align}
  V^{(2)}_T=
  \frac{C_F\alpha _s}{4m^2r^3} S_{12},
  \quad\quad S_{12}
  =2\left(
  3\frac{(\vec{r}\cdot \vec{S})^2}{r^2}-\vec{S}^2
  \right) .
  \label{}
\end{align}
Its matrix element can be
factorized into two parts:
\begin{align}
  \bra{nlsj}V^{(2)}_T\ket{nl'sj}
  =\bra{nl}\frac{C_F \alpha_s}{4m^2r^3}\ket{nl'}
  \bra{lsj}S_{12}\ket{l'sj},
  \label{}
\end{align}
where the first factor represents the
radial part and
the second factor represents the
angular and spin part.

The radial matrix has a form
\begin{align}
\frac{C_F\alpha _s}{4m^2}
  \bra{nl}\frac{1}{r^3}\ket{nl'}
  =
    \begin{blockarray}{c@{}cccccc@{\hspace{4pt}}cr}
    \mLabel{l'=} & \mLabel{0} & \mLabel{1} & \mLabel{2} & \mLabel{3} & \mLabel{\cdots} & \mLabel{n-1} & & \\
    \begin{block}{(c@{\hspace{5pt}}cccccc@{\hspace{5pt}}c)r}
      &\star&\star&0& 0& \cdots & 0&& \mLabel{l=0~~~} \\
      &\star&\star&\star& 0& \cdots & 0&& \mLabel{1~~~} \\
      &0&\star&\star& \star & \cdots & 0&& \mLabel{2~~~} \\
      &0&0&\star&\star& \cdots &0&& \mLabel{3~~~} \\
      &\vdots &\vdots &\vdots &\vdots &\ddots &\vdots &&\\
      &0&0&0&0& \cdots &\star&& \mLabel{n-1} \\
    \end{block}
  \end{blockarray},
  \label{eq:radial}
\end{align}
where the star ($\star$) denotes a non-zero value.
(This can easily be  shown using the generating function for
the Laguerre polynomial.)
Namely, the radial matrix element vanishes in the case $| l-l' |\geq 2$.

On the other hand, 
because of the parity of $S_{12}$ and the orbital wave function, 
$S_{12}$ can mix the state of $l=j\pm 1, s=1$ with $l'=j\pm 1,s=1$,
and of $l=j, s=1$ with $l'=j, s=1$. Other matrix
elements vanish. By explicit computation the angular
matrix elements are given by
\cite{Ulehla1969, Kwong:1988gm, Cordon:2009pj}
\begin{align}
  &\bra{lsj}S_{12}\ket{l'sj}=
  \begin{blockarray}{c@{}ccc@{\hspace{4pt}}cl}
    & \mLabel{l'=j-1} & \mLabel{j} & \mLabel{j+1} & & \\
    \begin{block}{(c@{\hspace{5pt}}ccc@{\hspace{5pt}}c)l}
      &-\frac{2(j-1)}{2j+1}&0&\frac{6\sqrt{j(j+1)}}{2j+1} & & \mLabel{l=j-1} \\
      & 0 & 2 & 0 & & \mLabel{l=j} \\
      &\frac{6\sqrt{j(j+1)}}{2j+1} &0&-\frac{2(j+2)}{2j+1}& & \mLabel{l=j+1} \\
    \end{block}
  \end{blockarray}
~~~~~~
\mbox{for $s=1,j\geq 1$}, 
  \label{eq:angular}
\\
 & \bra{lsj}S_{12}\ket{l'sj}=-4 
~~~~~~~~~~~~~~~~~~~~~~~~~~~~~~~~~~~~~
\mbox{for $l=l'=1,s=1,j=0$},
\end{align}
and $\bra{lsj}S_{12}\ket{l'sj}=0$ otherwise.
Thus, the only non-vanishing off-diagonal matrix elements are
the ones with $| l-l' |=2$.

Combining eqs.\eqref{eq:radial}, \eqref{eq:angular},
we obtain
\begin{align}
  \bra{nlsj}V^{(2)}_T\ket{nl'sj}
  =\bra{nl}\frac{C_F\alpha_s}{4m^2r^3}\ket{nl'}
  \bra{lsj}S_{12}\ket{l'sj}=0
  \label{}
\end{align}
for all $n,l,l'(\neq l),s,j$.

\section{Enhanced corrections at higher orders}
\label{s4}

The analysis of the previous section
does not apply to the tensor operator in the
third-order potential,
\begin{align}
V^{(3)}_T=&
\frac{C_F\alpha_s}{2m^2r^3}
\frac{\alpha_s}{\pi}
\left[
\frac{1}{72}\Bigl\{ C_A(97+18 L_m-18 L_r)
+4(9C_F-5T_Fn_l)\Bigr\}
+\frac{1}{24}\beta _0(3L_r-8)
\right]
S_{12},
\label{}
\end{align}
where
$L_m=\log (\mu ^2/m^2)$,
$L_r=\log (e^{2\gamma _E} \mu ^2r^2)$.
The essential difference between
$V^{(2)}_T$ and $V^{(3)}_T$
originates from the $\log r$ terms,
resulting in non-zero radial
matrix elements for $| l-l' |\geq 2$.
Indeed the off-diagonal matrix elements
of $\bra{nlsj}V^{(3)}_T\ket{nl'sj}$
have non-zero values.
It follows that the second order
correction to the wave function
$c^{(2)}$
in eq.\eqref{eq:c2}
has non-zero values, as well as
the cross-term of 
$\braket{V^{(3-i)}}$
and $d^{(i)}$ in eq.\eqref{eq:d3}
also has nonzero values.

Up to this point we considered enhanced contributions from
the intermediate states whose degeneracy is lifted by the
order $\varepsilon$ perturbation.
Let us comment on contributions from the states whose
degeneracy is lifted first at order $\varepsilon^2$,
namely, from the multiplets of fine and hyperfine
splittings [with the same $(n,l)$ but different $(s,j)$].
Such contributions, if they exist, would give rise
to more pronounced enhancement effects
than those considered so far,
since the level splittings which enter the denominator 
are order $\varepsilon^2$.
In fact such contributions are absent to all orders in $\varepsilon$, since
the transition matrix elements of $V^{(i)}$ between the states with
the same $(n,l)$ but different $(s,j)$ vanish
by parity and charge-parity conservation of QCD.\footnote{
The parity and charge-parity of the heavy quarkonium system
of the same flavor are given, respectively, by
$P=(-1)^{l+1}$ and $C=(-1)^{l+s}$
\cite{Lucha:1991vn}.
Hence, variations of $l$ and $s$, respectively, are allowed only by even
numbers.  
Since $s=0,1$, this means $s$ cannot change and $l$ can change
only by 0 or 2.
}
Thus, the order $\varepsilon^2$
quasi-degeneracy of the multiplets with the same $(n,l)$
does not give enhanced contributions to
either the energy level or the wave function.

Taking into account the fact that
$c^{(1)}=0,\ c^{(2)}\neq 0$ in eq.\eqref{eq:wavefunc},
the fourth- and fifth-order corrections
to the energy level are given by
\begin{align}
E^{(4)}_{nlsj}
&=\bra{nl}V^{(4)}\ket{nl}
+\sum _{i=1}^3
\sum _{n'\neq n,\ l'}
\frac{\bra{nl}V^{(4-i)}\ket{n'l'}d^{(i)}_{n'l';nl}}{E^{(0)}_{n}-E^{(0)}_{n'}}\,,
\label{eq:energy4}\\
E^{(5)}_{nlsj}
&=\bra{nl}V^{(5)}\ket{nl}
+\sum _{i=1}^4
\sum _{n'\neq n,\ l'}
 \frac{\bra{nl}V^{(5-i)}\ket{n'l'}d^{(i)}_{n'l';nl}}{E^{(0)}_{n}-E^{(0)}_{n'}}
+\sum_{l'\neq l}
\frac{\bra{nl}V^{(3)}\ket{nl'}c^{(2)}_{nl';nl}}{E^{(1)}_{nl}-E^{(1)}_{nl'}}\,.
\label{eq:energy5}
\end{align}
Note that the fourth-order correction
does not contain the enhanced corrections from
quasi-degeneracy,
which can be seen from the absence of
the $c$-term in eq.\eqref{eq:energy4}.

\section{Conclusions}

We have reconsidered the perturbation theory for
the Schr$\ddot{\mathrm{o}}$dinger equation
of the heavy quarkonium system in pNRQCD,
taking into account contributions of quasi-degenerate states.
As expected there are enhanced contributions
which rearrange the order counting of the expansion.
(In other words, it can be regarded as the question on
a proper order counting
of the $l$-changing mixing effects.)
At the (naive) lowest order,
the effect from the quasi-degenerate states
is induced only from one type of
off-diagonal matrix elements in the $l$-space,
$\bra{nlsj}V^{(2)}_T\ket{nl'sj}$.
This matrix element vanishes,
hence the quasi-degenerate correction
vanishes at the naive lowest order.
As a result this type of corrections are expected to appear first
at the fifth order in the energy level and
at the second order in the wave function.
The contribution to the heavy quark threshold cross
section or leptonic decay width is expected to start at the fourth order.

Thus, these specific corrections turn out to be irrelevant
with respect to the current highest-level perturbative QCD calculations
(energy levels and leptonic decay width at NNNLO).
We think that this fact itself should
be stated clearly.
It should also be noted that, since the enhanced corrections
to the wave function are expected to appear already at the second order,
they may be important for other physical observables, such as
level transition rates in certain channels.
We also note that,
even if the flavors of quark and antiquark are different (such
as the $B_c$ system), 
enhanced quasi-degenerate 
contributions to the NNNLO spectrum
\cite{Peset:2015vvi} ($l$-changing mixing)  vanish
similarly to the equal flavor case.\footnote{
In this case there are well-known mixing effects 
between different $s$ states from the second-order energy levels,
which correspond to the different diagonalizing
basis for $V^{(2)}$ for the same $(n,l)$.
}
These subjects will be discussed separately.

We remark that if one had a sufficiently wide knowledge one 
could reach the same conclusion without any computation,
since all the necessary results were available in the literature.
In this sense there is hardly any truly new ingredient in the
present work.
We find, however, that it is not easy to
collect these pieces of information together,
in particular since the old results on perturbative
computation of bound states before the advent of
modern EFT 
are scattered through literature in an unorganized way.
At least it would be meaningful to bring 
these results to the attention of 
experts at the forefront.

\section*{Appendix: General terms of perturbative series}

In this appendix,
we derive the general expressions for the
energy level correction
and the wave function correction
at the $N$-th order, in the case that the degeneracy is lifted
by the first-order and second-order corrections stepwise.
The expressions shown in Secs.~\ref{sec:2} and \ref{s4} are
special cases of the general expressions
eqs.\eqref{app:energy},
\eqref{app:d2} and
\eqref{app:c2}.

We expand
the Schr$\ddot{\mathrm{o}}$dinger
equation as
\begin{align}
  \left(
  H^{(0)}+\sum_{i=1}^{\infty}\varepsilon^i
  V^{(i)}
  \right)
  \left(
  \sum_{i'=0}^{\infty}
  \varepsilon^{i'}
  \Ket{\Psi_{nlsj}^{(i')}}
  \right)
  =\left(
  \sum_{i=0}^{\infty}
  \varepsilon^i
  E_{nlsj}^{(i)}
  \right)
  \left(
  \sum_{i'=0}^{\infty}
  \varepsilon^{i'}
  \Ket{\Psi_{nlsj}^{(i')}}
  \right).
  \label{app:sch}
\end{align}
The labels $s$ and $j$
are suppressed
in the rest of this appendix.
For notational simplicity,
we redefine the
wave function corrections as
$\tilde{c}^{(i)}_{nl';nl}=c^{(i)}_{nl';nl}/(E^{(1)}_{nl}-E^{(1)}_{nl'})$
and
$\tilde{d}^{(i)}_{n'l';nl}=d^{(i)}_{n'l';nl}/(E^{(0)}_{n}-E^{(0)}_{n'})$.
Then
the full wave function is given by
\begin{align}
  \Ket{\Psi_{nl}}&=
  \sum_{i=0}^{\infty}
  \varepsilon^{i}
  \Ket{\Psi_{nl}^{(i)}}
  =
  \ket{nl}
  +\sum _{i=1} ^\infty
    \sum_{l'\neq l}^{}
  \varepsilon^i
    \ket{nl'}
    \tilde{c}^{(i)}_{nl';nl}
  +\sum _{i=1}^\infty 
    \sum_{n'\neq n,\ l'}^{}
  \varepsilon^i
    \ket{n'l'}
    \tilde{d}^{(i)}_{n'l';nl}.
  \label{app:wavefunc}
\end{align}
In addition,
we write
$V^{(i)}_{n'l';nl}=\bra{n'l'}V^{(i)}\ket{nl}$
in the following.

The coefficient of $\varepsilon^N$ of
the Schr$\ddot{\mathrm{o}}$dinger
equation~\eqref{app:sch} reads
\begin{align}
  V^{(N)}\ket{nl}
  +\sum_{i=1}^{N-1}
  V^{(N-i)}\Ket{\Psi^{(i)}_{nl}}
  +H^{(0)}\Ket{\Psi^{(N)}_{nl}}
  =E^{(N)}_{nl}\ket{nl}
  +\sum_{i=1}^{N-1}
  E^{(N-i)}_{nl}\Ket{\Psi^{(i)}_{nl}}
  +E^{(0)}_{n}\Ket{\Psi^{(N)}_{nl}}.
  \label{app:eq}
\end{align}
The correction to the energy level is obtained by
multiplying eq.\eqref{app:eq} by $\bra{nl}$ from the left.
Since $\Braket{nl|\Psi ^{(i)}_{nl}}=0$ for $i\geq 1$,
the only remaining term on
the right-hand side 
is $E^{(N)}_{nl}$.
Then we obtain
\begin{align}
  E^{(N)}_{nl}
  =V^{(N)}_{nl;nl}
  +\sum_{i=1}^{N-1}
  \sum_{l''\neq l}^{}
  V^{(N-i)}_{nl;nl''}
  \tilde{c}^{(i)}_{nl'';nl}
  +\sum_{i=1}^{N-1}
  \sum_{n''\neq n,\ l''}^{}
  V^{(N-i)}_{nl;n''l''}
  \tilde{d}^{(i)}_{n''l'';nl}.
  \label{app:energy}
\end{align}

Next we consider the correction
to the wave function.
Multiplying eq.\eqref{app:eq} by $\bra{n'l'}$ from the left,
we obtain
\begin{align}
  V^{(N)}_{n'l';nl}
  +\sum_{i=1}^{N-1}
  \sum_{l''\neq l}^{}
  V^{(N-i)}_{n'l';nl''}
  \tilde{c}^{(i)}_{nl'';nl}
  +\sum_{i=1}^{N-1}
  \sum_{n''\neq n,\ l''}^{}
  V^{(N-i)}_{n'l';n''l''}
  \tilde{d}^{(i)}_{n''l'';nl}
  +E^{(0)}_{n'}
  \tilde{d}^{(N)}_{n'l';nl}
  ~~~~~~~~
  \nonumber\\
  =\sum_{i=1}^{N-1}
  E^{(N-i)}_{nl}
  \tilde{d}^{(i)}_{n'l';nl}
  +E^{(0)}_{n}
  \tilde{d}^{(N)}_{n'l';nl},
  \label{app:d}
\end{align}
where $\tilde{d}^{(N)}_{n'l';nl}$ is separated outside
the summation.
Solving eq.\eqref{app:d} for
$d^{(N)}_{n'l';nl}$,
we obtain
\begin{align}
  d^{(N)}_{n'l';nl}
  =  V^{(N)}_{n'l';nl}
  +\sum_{i=1}^{N-1}
  \sum_{l''\neq l}^{}
  V^{(N-i)}_{n'l';nl''}
  \tilde{c}^{(i)}_{nl'';nl}
  +\sum_{i=1}^{N-1}
  \sum_{n''\neq n,\ l''}^{}
  V^{(N-i)}_{n'l';n''l''}
  \tilde{d}^{(i)}_{n''l'';nl}
  -\sum_{i=1}^{N-1}
  E^{(N-i)}_{nl}
  \tilde{d}^{(i)}_{n'l';nl}.
  \label{app:d2}
\end{align}
Note that the left-hand-side of
eq.~\eqref{app:d2} is
not $\tilde{d}^{(N)}_{n'l';nl}$
but $d^{(N)}_{n'l';nl}$.

The case with $\tilde{c}^{(N)}_{nl';nl}$
is similar to that of $\tilde{d}^{(N)}_{n'l';nl}$
except that we first need to
replace $N\to N+1$ in eq.\eqref{app:eq}.
Then multiplying it by $\bra{nl'}$ from the left,
we obtain
\begin{align}
  V^{(N+1)}_{nl';nl}
  +\sum_{i=1}^{N-1}
  \sum_{l''\neq l}^{}
  V^{(N-i+1)}_{nl';nl''}
  \tilde{c}^{(i)}_{nl'';nl}
  +E^{(1)}_{nl'}
  \tilde{c}^{(N)}_{nl';nl}
   +\sum_{i=1}^{N}
  \sum_{n''\neq n,\ l''}^{}
  V^{(N-i+1)}_{nl';n''l''}
  \tilde{d}^{(i)}_{n''l'';nl}
  ~~~~~~~~
  \nonumber\\
  =\sum_{i=1}^{N-1}
  E^{(N-i+1)}_{nl}
  \tilde{c}^{(i)}_{nl';nl}
  +  E^{(1)}_{nl}
  \tilde{c}^{(N)}_{nl';nl}.
  \label{app:c}
\end{align}
Solving 
for $c^{(N)}_{nl';nl}$,
we obtain
\begin{align}
  c^{(N)}_{nl';nl}
  =V^{(N+1)}_{nl';nl}
  +\sum_{i=1}^{N-1}
  \sum_{l''\neq l}^{}
  V^{(N-i+1)}_{nl';nl''}
  \tilde{c}^{(i)}_{nl'';nl}
   +\sum_{i=1}^{N}
  \sum_{n''\neq n,\ l''}^{}
  V^{(N-i+1)}_{nl';n''l''}
  \tilde{d}^{(i)}_{n''l'';nl}
  -\sum_{i=1}^{N-1}
  E^{(N-i+1)}_{nl}
  \tilde{c}^{(i)}_{nl';nl}.
  \label{app:c2}
\end{align}
Again we note the difference between $c$ and $\tilde{c},\tilde{d}$
on both sides.

Thus, we obtain
the energy level correction~\eqref{app:energy}
and 
wave function correction~\eqref{app:d2},
\eqref{app:c2} at the $N$-th order
in a recursive form.

\section*{Acknowledgements}

The authors are grateful to Jorge Segovia
for bringing their attention to the
corrections from quasi-degenerate states.
The works of Y.K.\ and Y.S., respectively,
were supported in part by Grant-in-Aid for
scientific research Nos.\ 26400255 and 26400238 from
MEXT, Japan.

\end{document}